\documentclass[final,twocolumn,10pt,a4paper]{elsarticle}

\usepackage{amssymb}
\usepackage{hyperref}
\usepackage{algorithm}
\usepackage{algorithmicx}
\usepackage{algpseudocode}
\usepackage{amsmath}

\usepackage[nolist,nohyperlinks]{acronym}

\usepackage{tabularx}
\usepackage{booktabs}

\usepackage{listings}
\usepackage{xcolor}
\usepackage{soul}
\usepackage{ulem}
\usepackage{rotating}
\usepackage{enumitem}
\usepackage[margin=2cm]{geometry}

\sethlcolor{yellow}

\definecolor{codegreen}{rgb}{0,0.6,0}
\definecolor{codegray}{rgb}{0.5,0.5,0.5}
\definecolor{codepurple}{rgb}{0.58,0,0.82}
\definecolor{backcolour}{rgb}{0.95,0.95,0.92}
\definecolor{owngreen}{rgb}{0.0, 0.0, 1.0} 

\lstdefinestyle{mystyle}{
    backgroundcolor=\color{backcolour},   
    commentstyle=\color{codegreen},
    keywordstyle=\color{magenta},
    numberstyle=\tiny\color{codegray},
    stringstyle=\color{codepurple},
    basicstyle=\ttfamily\footnotesize,
    breakatwhitespace=false,         
    breaklines=true,                 
    captionpos=b,                    
    keepspaces=true,                 
    numbers=left,                    
    numbersep=5pt,                  
    showspaces=false,                
    showstringspaces=false,
    showtabs=false,                  
    tabsize=2,
    escapeinside={法}{法}
}

\journal{SoftwareX}

\begin{document}
\renewcommand{\labelenumii}{\arabic{enumi}.\arabic{enumii}}

\begin{frontmatter}



\title{Semi-asynchronous Federated Learning in Flower: Framework Extension and Performance Assessment}


\author[inst1]{Víctor Hidalgo-Izquierdo\corref{cor1}}
\ead{victor.hidalgo@uclm.es}
\author[inst2]{Carmen Carrión\fnref{fn1}}
\author[inst2]{Blanca Caminero\fnref{fn1}}

\cortext[cor1]{Corresponding author}
\fntext[fn1]{All other authors contributed equally to this work}

\address[inst1]{Instituto de Investigación en Informática de Albacete, Universidad de Castilla-La Mancha, Investigación 2, Albacete, 02071, Castilla-La Mancha, Spain}
\address[inst2]{Departamento de Sistemas Informáticos, Universidad de Castilla-La Mancha, Edificio Juan Manuel, Campus Universitario, Albacete, 02071, Castilla-La Mancha, Spain}

\begin{abstract}
\textit{This paper presents an extension of the Flower federated learning framework to support \ac{SAFL}. The proposed approach adapts the traditional synchronous paradigm to better handle client heterogeneity and straggler effects. By introducing a semi-asynchronous training strategy, the system allows partial synchronization among clients while maintaining training efficiency and scalability. We implement and evaluate the proposed modification within Flower, instantiated as the FedSaSync strategy, demonstrating improved robustness and reduced idle time compared to fully synchronous baselines in heterogeneous environments. The results show that \ac{SAFL} can balance convergence stability and system efficiency in heterogeneous environments typical of edge and distributed learning scenarios.}

\end{abstract}


\end{frontmatter}


\section*{Metadata}
The metadata of this project is listed in Table~\ref{tab:code_metadata}.
\label{sec:metadata}

\begin{table*}[h]
\centering
\small
\begin{tabularx}{\textwidth}{cXX}
\toprule
\textbf{Nr.} & \textbf{Code metadata description} & \textbf{Metadata} \\
\midrule
C1 & Current code version & v1.2 \\
C2 & Permanent GitHub link to code/repository used for this code version & \url{https://github.com/VictorHidalgoUCLM/SAFlwr/tree/main/baselines/fedsasync}\\
C3 & Legal Code License & Apache-2.0 license \\
C4 & Code versioning system used & Git \\
C5 & Software code languages, tools, and services used & Python, Flower, PyTorch \\
C6 & Compilation requirements, operating environments \& dependencies & Requirements: Python environment with Python 3.12.12, Dependencies: flwr[simulation] $\geq$ 1.24.0, flwr-datasets[vision] $\geq$ 0.5.0, torch $=$ 2.8.0, torchvision $=$ 0.23.0 \\
C7 & If available Link to developer documentation/manual & \url{https://github.com/flwrlabs/flower} \\
C8 & Support email for questions & victor.hidalgo@uclm.es \\
\bottomrule
\end{tabularx}
\caption{Code metadata}
\label{tab:code_metadata} 
\end{table*}

\section{Motivation and significance}
\label{sec:motivation}
The rapid growth of distributed data-generating devices has led to large-scale data availability across edge environments, enabling new opportunities for data-driven analysis and decision-making~\cite{Liu2023IoT}.

Traditional \ac{ML} approaches rely on centralized cloud infrastructure, in which data collected by edge devices is transmitted to remote servers for training. However, this paradigm raises significant concerns about communication overhead, latency, privacy, and security, thereby motivating the development of decentralized learning approaches~\cite{Andriulo2024}.

In this context, \ac{FL} has emerged as a promising solution to these limitations by shifting the learning process closer to the data sources~\cite{alsharif_contemporary_2024, neto_survey_2023, almanifi_communication_2023, houidi_federated_2024, boobalan_fusion_2022}. \ac{FL} follows a client-server architecture in which a central server orchestrates the training process, and multiple clients perform local model training on their private datasets. Clients periodically transmit model updates to the server, which aggregates them into a global model and redistributes the updated parameters. By avoiding direct data sharing, \ac{FL} significantly mitigates privacy and security concerns while reducing communication latency and bandwidth consumption.

However, conventional \ac{FL} typically relies on a synchronous training process, where the global model is updated only after receiving contributions from all participating clients~\cite{abreha_federated_2022}. In heterogeneous \ac{IoT} environments, this rigid synchronization requirement becomes a major limitation, as slower or less capable devices, often referred to as \textit{stragglers}, can delay the entire training process~\cite{he_fairness-guaranteed_2025}. Consequently, the overall learning pace is effectively determined by the slowest participant, reducing system efficiency and scalability.

\ac{AFL} partially addresses these problems by loosening the strict synchronization requirement, enabling clients to submit updates independently. Nonetheless, this flexibility may reduce efficiency and hinder model convergence, especially in resource-limited settings where outdated client updates can seriously affect convergence. These issues justify adopting \ac{SAFL}, which strikes a balance between full synchronization and asynchrony. \ac{SAFL} has recently received increasing attention in the literature and is designed to address robustness and energy efficiency in heterogeneous edge environments~\cite{xu_asynchronous_2023}.

Several open-source frameworks have been proposed to facilitate research and development in \ac{FL}, including TensorFlow Federated (TFF)~\cite{TFF}, Federated AI Technology Enabler (FATE)~\cite{FATE}, Flower~\cite{Flower_framework}, or PySyft~\cite{PySyft}. Among these, Flower stands out for its modular architecture, lightweight design, active community, and ongoing development. Unlike ecosystem-dependent alternatives, Flower decouples the federated infrastructure from the underlying local training logic, allowing seamless integration with any \ac{ML} framework (e.g., PyTorch, TensorFlow, or JAX). This modular architecture, backed by an active community and continuous development, provides the ideal flexibility for implementing and benchmarking custom, community-driven training and aggregation strategies~\cite{flower_baselines}.

Several studies have investigated semi-asynchronous synchronization methods, each with different formulations. A typical variant introduces semi-asynchronous training mechanisms that control the degree of client synchronization, as in FedSA~\cite{Ma2021-FedSA}, ASAFL~\cite{Chen_ASAFL}, FedSAP~\cite{zhao_FedSAP}, ASFL~\cite{yu_asfl_2023}, or SASAFL~\cite{Yu2024-SASAFL}.

A major challenge in current \ac{FL} research is the lack of a standardized framework for synchronization strategies, which often forces researchers to rely on custom, isolated simulation environments, severely hindering the reproducibility of results. While Flower has established itself as a mature, production-ready framework capable of becoming the industry standard, it currently lacks support for semi-asynchronous mechanisms. To address this limitation, this work implements the core semi-asynchronous orchestration at the application level within Flower. By doing so, we provide an accessible and standardized environment that enables future \ac{SAFL} research to be implemented, benchmarked, and reproduced directly on a reliable platform.

To the best of our knowledge, this is the first implementation of a semi-asynchronous training method within the Flower framework. The software presented in this work represents a valuable contribution to the Flower ecosystem. Although \ac{SAFL} has attracted growing interest in recent years, support for this synchronization paradigm is still limited in widely adopted \ac{FL} frameworks.

In summary, the primary contributions of FedSaSync are outlined below:

\begin{itemize}[leftmargin=1.2em, nosep]
    \item \textbf{Semi-Asynchronous Aggregation Mechanism extension:} This work introduces a semi-asynchronous aggregation method for \ac{FL}, implemented as the FedSaSync strategy. Global model updates are triggered once a predefined number of clients, $M$, have completed their local training, decoupling aggregation from full client synchronization.
    \item \textbf{Support for Heterogeneous and Time-Varying Clients:} The proposed mechanism enables a faithful realization of semi-asynchronous behavior in environments with heterogeneous and time-varying client execution times, reducing the impact of slow participants on the aggregation process.
    \item \textbf{Open-Source Implementation for Reproducible Research:} This work provides an open and detailed implementation of the proposed \ac{SAFL} mechanism. Following the architectural design principles of the Flower baselines, the software serves as a reusable and extensible foundation for future research and reproducible experimentation.
    \item \textbf{Systematic Evaluation of Semi-Asynchronous Training:} The proposed extension enables a systematic study of the effects of different semi-asynchronous configurations, facilitating analysis of convergence, training efficiency, and client participation dynamics across varying levels of heterogeneity.
\end{itemize}

To explicitly clarify its architectural positioning, \texttt{FedSaSync} is designed as an open-source execution and orchestration middleware rather than a rigid, closed-form algorithmic solution. By strictly separating low-level concurrency orchestration (e.g., dynamic delays, task scheduling, and real-time observability metrics) from high-level algorithmic logic, \texttt{FedSaSync} provides a flexible, reproducible substrate for the FL community. Under this modular paradigm, specific algorithmic mechanisms—such as custom staleness control strategies (e.g., decay functions or thresholding, like FedMDS~\cite{zhang_fedmds_2023}), heterogeneity-aware optimization approaches (e.g., HA-HEFL~\cite{li_heterogeneity-aware_2026} or AMP-FL~\cite{chen_adaptive_2024}), security protocols (e.g., SSAA~\cite{li_ssaa_2026}), or concrete asynchronous variants (e.g., FedSA~\cite{Ma2021-FedSA}, ASAFL~\cite{Chen_ASAFL})—are treated as orthogonal policies that can be seamlessly implemented and benchmarked on top of the underlying orchestration infrastructure.

\section{Software description}
In this section, we present the software implementation that supports the semi-asynchronous \textit{FedSaSync} strategy within the Flower framework. First, we describe the overall software architecture of Flower through two class diagrams and the project directory structure, providing a high-level overview of the system and highlighting the components that were modified or extended. Subsequently, we detail the specific functionalities added or modified in our implementation, illustrated by the algorithm that formalizes and clarifies the proposed approach. Finally, auxiliary framework extensions developed to ensure determinism, monitoring, and overall utility during experimental evaluation are detailed.

\subsection{Software Architecture}
The software architecture follows the design principles of the Flower \ac{FL} framework. Flower is designed to enable a seamless transition from experimental research in simulation environments to system-level research on edge devices. Its main design goals include scalability, communication, and client agnosticity, privacy preservation, and flexibility to support both experimental studies and real-world deployments. Conceptually, Flower organizes \ac{FL} as the interaction between global and local computations, providing the infrastructure required to execute \ac{FL} processes at scale.

Our implementation extends the baseline Flower framework while preserving its modular architecture. To enable \ac{SAFL}, the functions responsible for synchronous coordination in the original implementation have been adapted while maintaining compatibility with the framework design. In the architectural diagrams presented in the following sections, the modified functions are highlighted in bold to distinguish them from the original Flower implementation.

\begin{figure*}[htbp]
    \centering
    \includegraphics[width=0.95\textwidth]{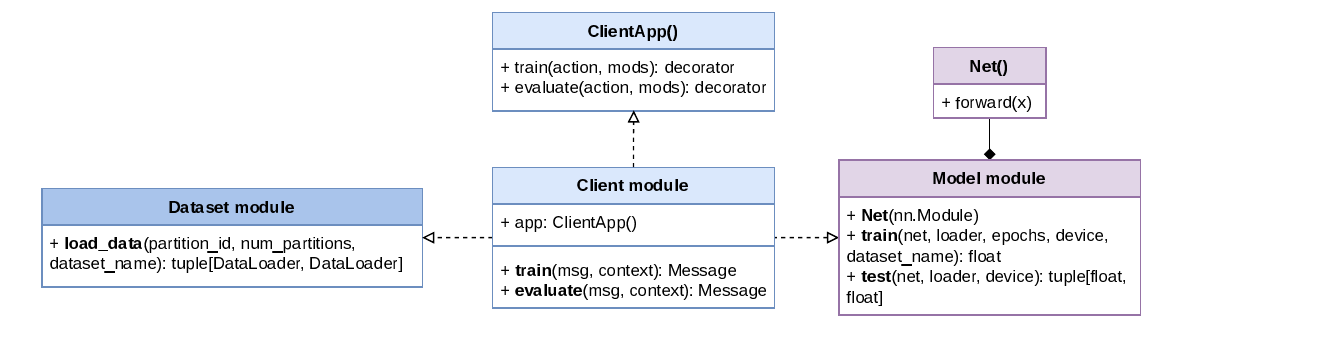}
    \caption{FedSaSync client class diagram.}
    \label{fig:fl_client}
\end{figure*}

The client-side architecture is illustrated in Figure~\ref{fig:fl_client}, where the \textbf{Client module} serves as the main component defining the behavior of a federated client. Its role is to encapsulate the local training and evaluation logic executed on each client. This module instantiates a \texttt{ClientApp()} object named \texttt{app} obtained from the Flower framework. The application exposes the \texttt{train} and \texttt{evaluate} functions through decorators, which can be overridden within the \textbf{Client module} to implement custom client-side logic.

In addition, the \textbf{Dataset module} and the \textbf{Model module} define the data and model components of the system, respectively. The former is responsible for generating the data loaders required by each client, while the latter defines the neural network architecture (\texttt{Net}) along with its corresponding training and evaluation procedures.

\begin{figure*}[htbp]
    \centering
    \includegraphics[width=0.90\textwidth]{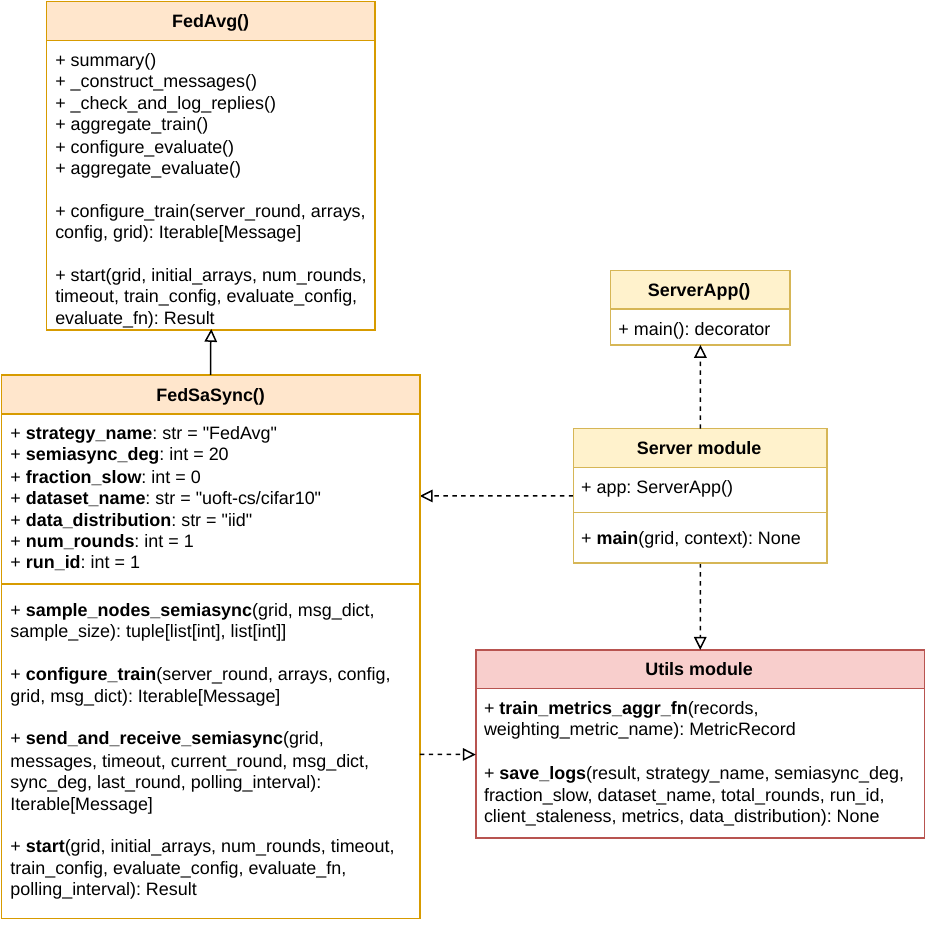}    \caption{FedSaSync server class diagram.}
    \label{fig:fl_server}
\end{figure*}

The server-side architecture, illustrated in Figure~\ref{fig:fl_server}, follows a similar design, in which the \textbf{Server module} is the key component that defines the behavior of the federated server. Its role is to coordinate the global training process and manage the aggregation of client updates. This module instantiates a \texttt{ServerApp()} object named \texttt{app}, also obtained from the Flower framework. The application exposes the \texttt{main} function via a decorator that selects and configures the \ac{FL} strategy used during training.

To support \ac{SAFL}, the server module implements a custom extension built on top of the Flower framework. This extension overrides the standard \texttt{FedAvg()} baseline and introduces the necessary components to enable semi-asynchronous coordination between clients and the server. Within this extended framework, we instantiate a semi-asynchronous strategy, \texttt{FedSaSync()}, which leverages \texttt{send\_and\_receive\_semiasync()} as the core mechanism governing synchronization and communication. This design provides a general interface for semi-asynchronous execution, allowing different aggregation behaviors to be implemented on top of the same underlying communication logic.

The \texttt{main()} function of the server module initializes and activates the \texttt{FedSaSync()} strategy instead of the default \texttt{FedAvg()} strategy (see Listing~\ref{lst:fedsa}), where the specific parameters added for our proposed strategy, namely \texttt{strategy\_name}, \texttt{semiasync\_deg}, \texttt{fraction\_slow}, and \texttt{dataset\_name}, are highlighted in blue and will be detailed in later sections. Additionally, both the \texttt{FedSaSync()} implementation and the server module rely on auxiliary utilities provided in the \textbf{Utils module}, enabling customized metrics aggregation and system logging. These additions are essential for monitoring training dynamics and collecting experimental data for subsequent analysis.

\begin{lstlisting}[language=Python, caption={main(): Initialization of FedSaSync strategy}, label={lst:fedsa}]
# Initialize FedSaSync strategy
strategy = FedSaSync(
    fraction_train=fraction_train,
    fraction_evaluate=fraction_evaluate,
    min_available_nodes=2,
    法{\color{owngreen}strategy\_name=strategy\_name,}法
    法{\color{owngreen}semiasync\_deg=semiasync\_deg,}法
    法{\color{owngreen}fraction\_slow=fraction\_slow,}法
    法{\color{owngreen}dataset\_name=dataset\_name,}法
    train_metrics_aggr_fn=train_metrics_aggr_fn,
)
\end{lstlisting}

\begin{figure*}[htbp]
    \centering
    \includegraphics[width=0.80\textwidth]{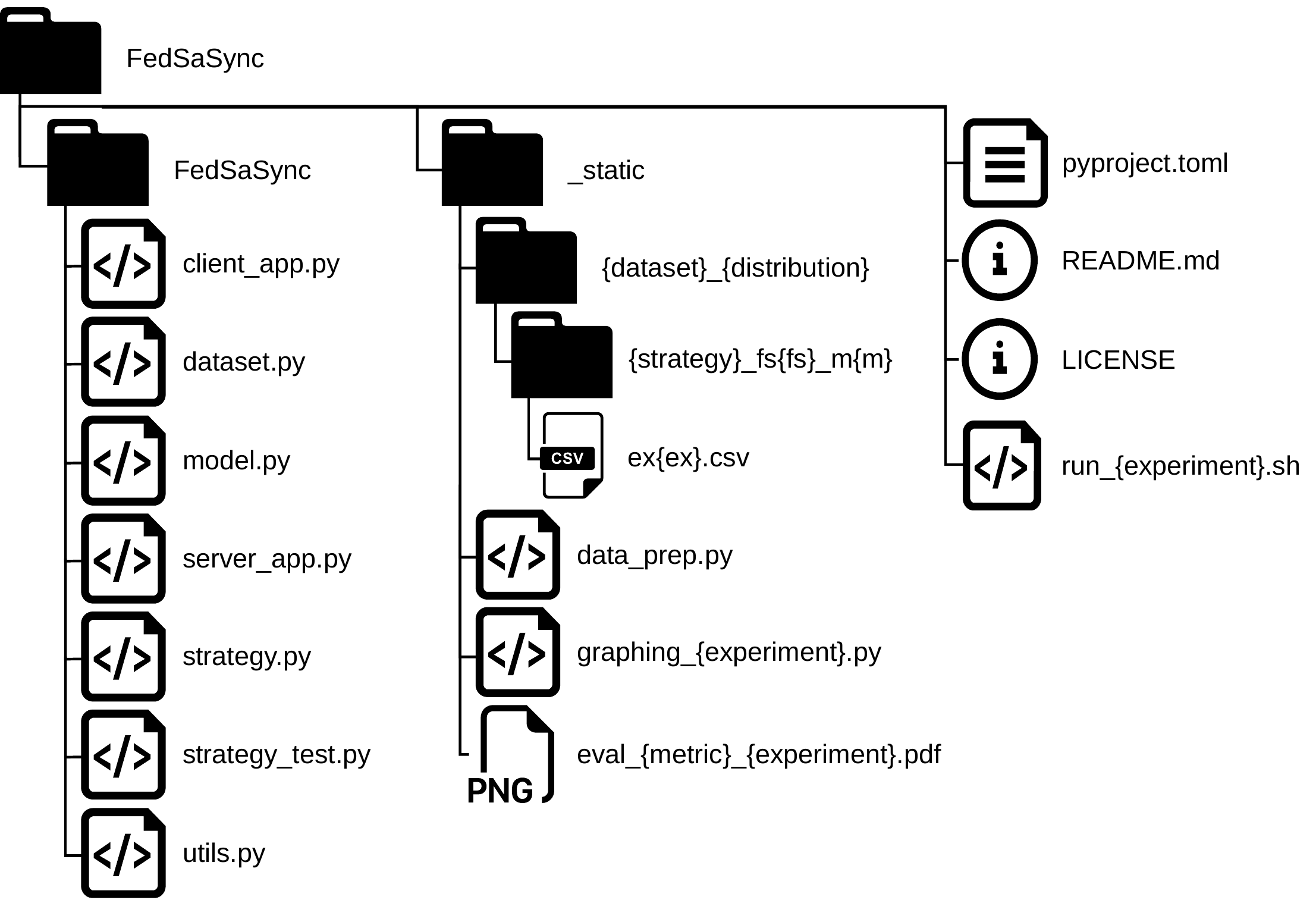}
    \caption{FedSaSync project architecture.}
    \label{fig:fl_architecture}
\end{figure*}

Finally, Figure~\ref{fig:fl_architecture} presents the core project structure for the base \texttt{FedSaSync} workflow. The \texttt{FedSaSync} directory contains all source code related to the client and server modules described above (alternative strategy adaptations, such as \texttt{FedProxSaSync}, follow an identical layout within the dedicated repository directories). Additionally, dedicated tests integrated alongside the strategy code (\texttt{fedsasync/strategy\_test.py}) ensure the correctness and reliability of both client and server components under timeout and failure conditions. The verification suite is executed from the \texttt{/fedsasync} root directory using \texttt{pytest}, yielding passing outcomes that confirm the coherence of semi-asynchronous coordination and staleness registration. The \texttt{pyproject.toml} file defines the project configuration and dependencies, which are detailed in subsequent sections. The \texttt{\_static} directory stores execution results in CSV format, while a dedicated Python script generates plots and saves the resulting figures in the same directory. Finally, the Bash scripts define and automate the experimental setup used in this work.

These components and data collection mechanisms form the foundation of our implementation. The following sections provide a detailed description of the specific functionalities implemented within this architecture.

\subsection{Software functionalities}
The main extension introduced to the Flower framework is a module that enables semi-asynchronous execution under different \ac{FL} strategies. As a core instantiation of this mechanism, we propose \texttt{FedSaSync}, which extends the synchronous \texttt{FedAvg} algorithm by supporting semi-asynchronous aggregation. In this setup, model updates are aggregated whenever a predefined number of clients complete their local training.

The implementation of the semi-asynchronous aggregation system follows the workflow described in Algorithm~\ref{alg:semi-asynchronous-send-receive}. The server keeps track of the clients currently involved in training and continuously collects their updates as they become available. Whenever the number of received updates reaches the semi-asynchronous synchronization degree $M$, an aggregation event is triggered without waiting for the remaining clients. This stands in sharp contrast to the traditional FedAvg strategy, which operates in a strictly synchronous manner and requires updates from all selected clients before triggering an aggregation, making it highly susceptible to the straggler problem. The clients that contributed to the aggregation are then released and become eligible for subsequent training tasks, making them available for assignment in subsequent rounds. To maintain round semantics, all rounds operate in a semi-asynchronous manner, except for the final round, which aggregates synchronously by waiting for all remaining client updates before ending the training process.

It is important to note that $M$ acts as a lower bound rather than a strict cardinality constraint. In practice, multiple clients may complete their local training simultaneously, leading to aggregation events involving more than $M$ updates when those updates become available in the same polling iteration. This behavior is intentional and allows the system to naturally exploit concurrent completions without enforcing artificial synchronization barriers.

This behavior is implemented as follows. After dispatching the training tasks for the sampled clients and registering the corresponding busy clients into the \texttt{msg\_dict} Python dictionary (lines 1--8), the server continuously polls for client replies (lines 13--21). Aggregation is triggered as soon as $M$ updates are available (lines 17--21), interrupting the waiting loop and enabling intermediate aggregation without blocking on slower clients. In the final round of the federated training, the server instead waits until all pending replies have been received before proceeding (line 17). Finally, clients whose updates have been processed are removed from the set of busy nodes (lines 22--26), making them available for new training tasks within the same global round.

To support this behavior, several methods of the FedSaSync strategy interface were extended. In addition, \texttt{sample\_nodes\_semiasync()} implements deterministic selection of available clients from the set of free nodes, ensuring that only eligible clients participate in training rounds.

Regarding update staleness, FedSaSync aggregates all $M$ received updates unconditionally using the underlying base strategy (e.g., FedAvg), regardless of whether a client update was computed against the current global model or an earlier round. By default, the framework neither discards nor down-weights delayed updates. Instead, our implementation preserves training round identifiers within each message's metadata (\texttt{msg.metadata.group\_id}). This design decouples the synchronization mechanism from specific aggregation heuristics, establishing a transparent baseline while allowing developers to easily subclass the strategy and implement custom policies, such as staleness decay weighting or bounded staleness thresholds.

The remaining modifications are auxiliary to the strategy's execution. The \texttt{\_\_init\_\_()} method was extended to include the additional parameters required by the semi-asynchronous mechanism, while \texttt{configure\_train()} integrates the custom client-selection procedure into the training pipeline. Finally, \texttt{start()} was adapted to manage semi-asynchronous execution, measure elapsed time between aggregation events, and record runtime logs for later analysis.

\begin{algorithm}[]
\caption{Semi-asynchronous Send and Receive}
\label{alg:semi-asynchronous-send-receive}
\begin{algorithmic}[1]
\Require Grid $G$, messages $\mathcal{M}$, timeout $T$ (optional),
         message dictionary $\texttt{msg\_dict}$,
         synchronization degree $M$, last round flag $L$, polling time $t_{poll}$
\Ensure Returned messages $\mathcal{R}$

\State $\texttt{msg\_ids} \gets G.\texttt{push\_messages}(\mathcal{M})$ \hfill
\Statex

\If{$\texttt{msg\_dict}$ is None} \hfill
    \State $\texttt{msg\_dict} \gets \emptyset$
\EndIf
\For{each $(\texttt{msg\_id}, \texttt{msg})$ in $\texttt{msg\_ids} \times \mathcal{M}$} \hfill
    \State $n \gets \texttt{msg}.\texttt{metadata}.\texttt{dst\_node\_id}$
    \State $\texttt{msg\_dict}[n] \gets \texttt{msg\_id}$ \hfill
\EndFor

\Statex
\State \textbf{delete} $\mathcal{M}$
\State $\mathcal{A} \gets \texttt{set}(\texttt{msg\_dict}.\texttt{values()})$ \hfill
\State $\mathcal{R} \gets [\,]$ \hfill
\State $t_{\text{end}} \gets \texttt{time}() + (T \text{ if } T \neq \text{None} \text{ else } 0)$ \hfill

\While{$T = \text{None}$ \textbf{or} $\texttt{time}() < t_{\text{end}}$} \hfill
    \State $\mathcal{R}_{\text{new}} \gets G.\texttt{pull\_messages}(\mathcal{A})$ \hfill
    \State $\mathcal{R} \gets \mathcal{R} \cup \mathcal{R}_{\text{new}}$ \hfill
    \State $\mathcal{A} \gets \mathcal{A} \setminus \textsc{Replies}(\mathcal{R}_{\text{new}})$ \hfill

    \If{$( \neg L \land |\mathcal{R}| \ge M ) \lor ( L \land \mathcal{A} = \emptyset )$}
    \State \textbf{break}
    \EndIf
    \State $\texttt{sleep}(t_{poll})$ \hfill
\EndWhile

\Statex
\For{each $n$ in $\texttt{keys}(\texttt{msg\_dict})$} \hfill
    \If{$\texttt{msg\_dict}[n] \notin \mathcal{A}$}
        \State \textbf{delete} $\texttt{msg\_dict}[n]$ \hfill
    \EndIf
\EndFor

\State \Return $\mathcal{R}$ \hfill
\end{algorithmic}
\end{algorithm}

\subsection{Auxiliary Framework Extensions}
To support the experimental evaluation of the proposed strategy, several auxiliary extensions were also incorporated into the framework:

\begin{enumerate}[leftmargin=1.2em, nosep]
    \item The data management pipeline was extended to support selecting different datasets, configuring statistical data distributions (e.g., IID, non-IID), and deterministic partitioning across clients. Additionally, model initialization and server-side parameter generation were adapted to ensure reproducible executions via a seed-based approach combined with a configurable \texttt{run-id} parameter that explicitly tracks individual execution runs.
    \item The client implementation was extended with monitoring capabilities. In particular, each client records its local training time, which the server later aggregates to analyze the system's temporal behavior. Support for emulated slow clients was also incorporated to evaluate the impact of heterogeneous training speeds, and the communication/coordination loop was updated to incorporate a configurable \texttt{polling\_interval}.
    \item The utility layer was enhanced with mechanisms to aggregate training-related metrics and generate execution logs. These logs store information such as training times, aggregation events, and round progression, along with advanced metrics tracking update staleness, client participation rates, and scheduler overhead, facilitating the analysis and reproducibility of experimental results.
\end{enumerate}

Finally, the configuration file \texttt{pyproject.toml} was extended with additional parameters (see Listing~\ref{lst:pyproject}) required by the proposed system, including the experiment name, the degree of semi-asynchrony, the fraction of slow clients, the selected dataset, the selected data distribution, selected polling interval, and run-id for reproducibility. These parameters offer a flexible way to configure and reproduce various experimental scenarios, with values overwritten during execution.

\begin{lstlisting}[float=htbp, language=Python, caption={pyproject.toml default configuration}, label={lst:pyproject}, escapeinside={(*}{*)}]
# Custom config values accessible via `context.run_config`
[tool.flwr.app.config]
name = "FedSaSync" (*\textbf{\color{blue}\# Added}*)
num-server-rounds = 50
fraction-train = 1.0
fraction-evaluate = 1.0
local-epochs = 1
semiasync-deg = 10 (*\textbf{\color{blue}\# Added}*)
fraction-slow = 0.0 (*\textbf{\color{blue}\# Added}*)
dataset-name = 
  "uoft-cs/cifar10" (*\textbf{\color{blue}\# Added}*)
data-distribution = "iid" (*\textbf{\color{blue}\# Added}*)
polling-interval = 0.05 (*\textbf{\color{blue}\# Added}*)
run-id = 1 (*\textbf{\color{blue}\# Added}*)
\end{lstlisting}

\subsection{Middleware Reusability and Strategy Adaptation}
To ensure \texttt{FedSaSync} serves as a general-purpose platform rather than a monolithic implementation, its core architecture decouples orchestration primitives from algorithmic optimization. Within the Flower ecosystem, advanced optimization strategies (e.g., \texttt{FedProx}, \texttt{FedOpt}) inherit from the foundational \texttt{FedAvg} workflow. Consequently, adapting our middleware to alternative derivative strategies requires only minimal adjustments, leveraging core utilities like \texttt{sample\_nodes\_semiasync} and \texttt{send\_and\_receive\_semiasync}. Specifically, integrating a custom-parameter strategy (such as \texttt{FedProx} and its proximal term $\mu$) involves three straightforward engineering steps: (1) declaring the new hyperparameter (e.g., \texttt{proximal-mu}) in \texttt{pyproject.toml}; (2) updating the client-side training pipeline to compute the proximal regularization term during local optimization, along with adapting the training algorithm to that of the original \texttt{FedProx}; and (3) configuring the server strategy class to inherit from \texttt{FedProx} instead of \texttt{FedAvg}, which seamlessly substitutes standard aggregation with semi-asynchronous primitives while setting the \texttt{FedProxSaSync} identifier for experiment tracking and file logging. Steps (2) and (3) are illustrated in Listing~\ref{lst:fedprox_adapt}, while a fully functional implementation is available in the public repository (\texttt{fedsasync/FedProxSaSync}; see the \texttt{README.md}).

\begin{lstlisting}[language=Python, caption={Schematic adaptation of \texttt{FedSaSync} for a FedProx-derived strategy.}, label={lst:fedprox_adapt}, float=*t, mathescape=true]
# Code modification at server_app.py
@app.main()
def main(grid: Grid, context: Context):
  # Retrieve proximal-mu value from run_config
  $\textsf{\textbf{proximal\_mu = float(context.run\_config["proximal-mu"])}}$

  # Add proximal_mu value to strategy_name
  # e.g., "FedProxSaSync_0.0"
  $\textsf{\textbf{strategy\_name = f"\{strategy\_name\}\_\{proximal\_mu\}"}}$ 

  strategy = FedSaSync(
    ...,
    strategy_name=$\textsf{\textbf{strategy\_name}}$,
  )

# Code modification at strategy.py
class FedSaSyncProx($\textsf{\textbf{FedProx}}$):
  def __init__(self, ..., $\textsf{\textbf{proximal\_mu}}$: float = 0.0):
    super().__init__(...)
    $\textsf{\textbf{self.proximal\_mu = proximal\_mu}}$

# Code modification at client_app.py
@app.train()
def train(context: Context):
  # Retrieve hyperparameter directly from Flower's run configuration
  $\textsf{\textbf{prox\_mu = float(context.run\_config.get("proximal-mu", 0.0))}}$

  # Execute local training passing the proximal parameter
  $\textsf{\textbf{train\_loss = train\_fn(..., proximal\_mu=prox\_mu)}}$
\end{lstlisting}

This modular design allows researchers to readily plug in any Flower-compatible, FedAvg-derived strategy, inheriting robust semi-asynchronous timing, buffering, and state management out of the box.

Building upon this flexible foundation, our architecture enables the implementation of various state-of-the-art semi-asynchronous strategies. Depending on their underlying operational mechanism, their integration compatibility spans three distinct levels: theoretically supported native and telemetry-driven frameworks (e.g., FedSA, ASAFL, ASFL) whose server orchestration and telemetry-fed buffers natively match our architecture; partially supported gradient-centric frameworks (e.g., FedSAP) requiring custom aggregation logic for multi-model structures; and decentralized autonomous frameworks (e.g., SASAFL) which demand core architectural extensions to support client-initiated communication flows.

\section{Illustrative examples}
This section presents an empirical evaluation of the proposed \ac{SAFL} framework under controlled conditions across both homogeneous (IID) and heterogeneous (non-IID) data distributions. The goal is to provide an initial understanding of the system's behavior under varying degrees of synchronization, client execution delays, and data heterogeneity, rather than to establish definitive performance guarantees. To this end, we analyze the interplay between aggregation frequency, computational and statistical client heterogeneity, and overall training efficiency.

The experiments were conducted under the architectural and runtime constraints summarized in Table~\ref{tab:runtime_specs}.

\begin {table*}[h!]
  \centering
  \small
  \begin{tabularx}{\linewidth}{l X}
  \toprule
  \textbf{Component / Parameter} & \textbf{Configuration / Specification} \\
  \midrule
  OS / Kernel & Linux (Kernel 6.8.0-138-generic, x86\_64) \\
  Processor (CPU) & Intel Core i7-12700 (12 cores: 8 Performance, 4 Efficient) \\
  Hardware Capacity & 20 logical CPU threads (12 physical cores, Hyper-Threading enabled) \\
  \ac{FL} Framework & Flower v1.37.0 \\
  Execution Backend & Ray v2.55.1 \\
  Scheduling Levels & Ray (logical task slots) + Linux CFS (thread time-sharing) \\
  Slow Client Delay & 5.0 seconds per \texttt{local-epoch} \\
  Concurrency Constraint & Bounded to 1 client per logical CPU thread in the Ray cluster \\
  \bottomrule
  \end{tabularx}
  \caption{Architectural and runtime specifications for the simulation environment.}
  \label{tab:runtime_specs}
\end{table*}

\noindent As outlined in the table, Ray manages execution concurrency through logical resource slots (operating independently of low-level operating-system core affinity), restricting active parallelism per host to the capacity of the node (set to 20 concurrent clients in our baseline evaluation environment for simplicity and consistency). In distributed simulation loops, if straggling instances saturate these execution slots prematurely, backpressure propagates through the scheduling queue and inadvertently starves subsequent tasks, distorting true wall-clock measurements due to local CPU oversubscription. Consequently, bounding concurrency to match each node's logical core count was adopted as a controlled experimental design choice in our testbed to prevent resource contention from skewing wall-clock fidelity, whereas scaling to larger populations is achieved by extending the Ray cluster across additional physical nodes.

\subsection{Experimental setup}
The experimental evaluation considers an image classification task using two standard benchmark datasets: CIFAR-10 and MNIST. Both datasets contain 10 classes and are distributed across 20 clients using an either IID or non-IID partitioning strategy, enabling the evaluation of both homogeneous and heterogeneous data conditions. This configuration ensures controlled baseline comparisons across data distributions.

The baseline model is a simple convolutional neural network implemented in PyTorch, following the default architecture provided in the Flower framework. The model is adapted to each dataset in terms of input dimensionality and learning rate configuration.

For CIFAR-10, training is performed over 50 communication rounds, while MNIST is trained over 25 rounds. The learning rate is set to 0.01 for CIFAR-10 and 0.05 for MNIST, following standard configurations for these benchmarks.

Unless otherwise specified, the system operates with 20 clients in total, all of whom participate in every training round (fraction train of 1.0). Each client is allocated 1 CPU core and no GPU resources. The default aggregation strategy is \texttt{FedSaSync}, with a semi-asynchronous synchronization degree $M = 20$ (synchrony), a default polling interval $t_{poll} = 0.05\text{s}$, and no artificially induced slow clients.

To evaluate the impact of heterogeneity and semi-asynchronous execution, additional experimental configurations are summarized in Table~\ref{tab:experiment_configs}. These configurations vary along four main factors: (i) the dataset (CIFAR-10 or MNIST), (ii) the data distribution (IID or non-IID), (iii) the fraction of slow clients (ranging from $0.0$ to $0.5$), and (iv) the synchronization degree $M$, which takes values in $\{8, 10, 12, 14, 16, 18, 20\}$ under the proposed FedSaSync strategy, along with a fully synchronous baseline using standard \texttt{FedAvg} for comparison, and the polling interval $t_{poll}$. The number of communication rounds and learning rates are kept fixed for each dataset across all configurations. These experiments are implemented through the scripts \texttt{run\_\{dataset\}\_experiments.sh}.

\begin{table*}[!h]
\centering
\caption{Experiment configurations.}
\label{tab:experiment_configs}
\resizebox{0.95\textwidth}{!}{%
\begin{tabular}{lllllll}
\toprule
Dataset name & Data Distribution & Slow clients (frac.) & Semi-asynchronous degree & Polling interval $t_{poll}$ & Number of rounds & Learning Rate \\
\midrule
{CIFAR10, MNIST} & {IID, non-IID} & {0.0 -- 0.5} & {8, 10, 12, 14, 16, 18, 20, FedAvg} & {0.05s} & \begin{tabular}[c]{@{}l@{}}CIFAR10=50\\MNIST=25\end{tabular} & \begin{tabular}[c]{@{}l@{}}CIFAR10=0.01\\MNIST=0.05\end{tabular} \\
\bottomrule
\end{tabular}
}
\end{table*}

\subsection{Evaluation methodology}
The results are presented in three subfigures per dataset, each corresponding to a different proportion of slow clients (\texttt{fraction\_slows}). Within each subfigure, different values of $M$ are compared.

Performance is evaluated in terms of both test loss and test accuracy as a function of wall-clock training time rather than communication rounds. This choice is motivated by the system's semi-asynchronous nature, where communication rounds are not uniformly defined across clients and therefore may not provide a strictly comparable unit of progress. Consequently, time-based curves are used as a more consistent evaluation metric across configurations.

In addition, heatmaps are incorporated to analyze update staleness and client participation dynamics, complemented by comprehensive tables that detail scheduler and model performance metrics beyond what is captured by the aggregate curves.

\subsection{Results}
We evaluate the performance of the semi-asynchronous framework across convergence dynamics, fairness, staleness, and scheduler overhead. While our primary analysis focuses on the IID data distribution, we also examine the key differences and performance shifts introduced by non-IID settings.

Selected results for the experiments carried out with the CIFAR-10 dataset are presented and discussed next. Results for the MNIST testbed (as well as additional experiments for CIFAR-10) can be found in the FedSaSync Github repository.

\subsubsection{Convergence and Efficiency}
\begin{figure*}[t]
    \centering
    \includegraphics[width=1.0\linewidth]{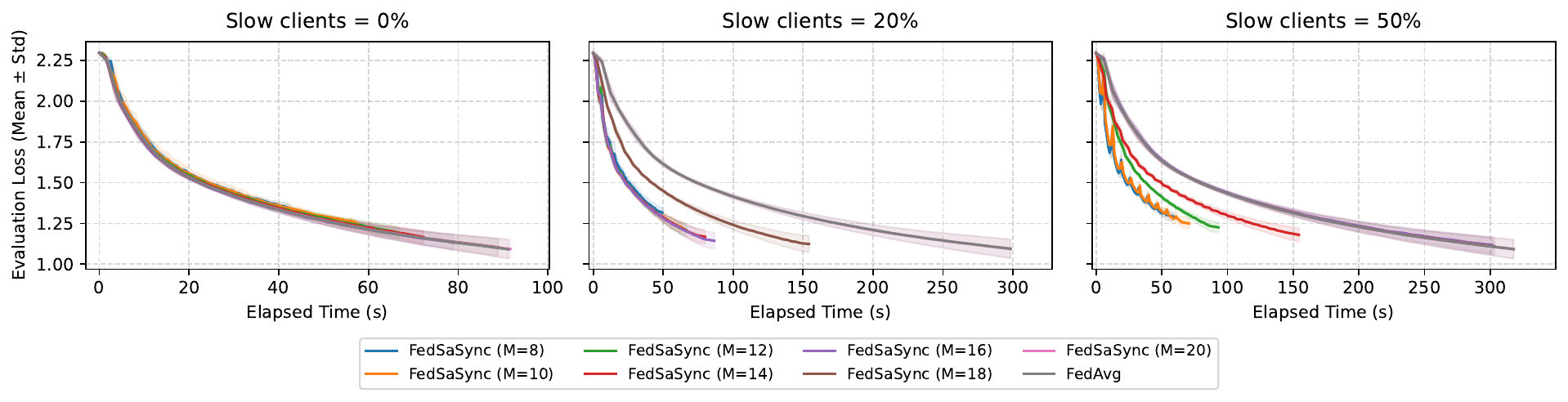}
    \caption{Evaluation loss versus wall-clock time for CIFAR-10 (IID) under different semi-asynchronous configurations and fractions of slow clients.}
    \label{fig:cifar_comparison_loss}
\end{figure*}

Figure~\ref{fig:cifar_comparison_loss} shows the training loss dynamics, confirming that our semi-asynchronous implementation naturally converges to standard synchronous \texttt{FedAvg} when $M = 20$. For smaller thresholds, system performance depends directly on whether $M$ bypasses stragglers: configurations where $M \leq N - N_{\text{slow}}$ operate at fast-client speeds, whereas higher $M$ values force the server to wait.

\begin{figure*}[t]
    \centering
    \includegraphics[width=1.0\linewidth]{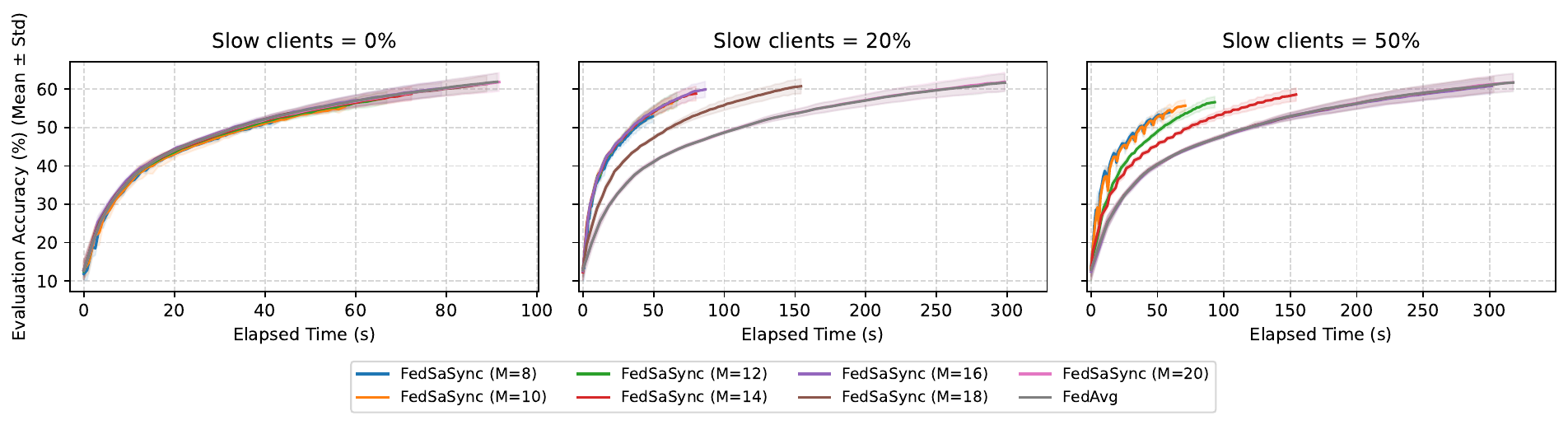}
    \caption{Evaluation accuracy versus wall-clock time for CIFAR-10 (IID) under different semi-asynchronous configurations and fractions of slow clients.}
    \label{fig:cifar_comparison_accuracy}
\end{figure*}

To complement the loss curves, Figure~5 illustrates test accuracy versus wall-clock time. Furthermore, Table~\ref{fig:cifar_comparison_accuracy} reports a comprehensive set of performance and efficiency metrics, exposing crucial operational trade-offs governed by the threshold $M$:

\begin{itemize}[leftmargin=1.2em, nosep]
  \item \textbf{Loss Reduction Efficiency ($\Delta \mathrm{loss} / \mathrm{second}$):} Favors lower $M$ configurations, as smaller thresholds allow global rounds to complete rapidly without waiting for stragglers.
  \item \textbf{Peak and Final Performance Ceiling (Max. Acc. \& Final Acc.)}: Consistently favors full synchrony or high $M$ settings, ensuring superior convergence stability and ultimate accuracy despite requiring more wall-clock time. Notably, in this specific experiment, the final accuracy mirrors the peak accuracy across all settings, as no late-stage degradation occurred during training.
  \item \textbf{Time-to-Accuracy (TTA) and Snapshots:} Lower $M$ values may miss stringent target thresholds entirely (yielding \textit{N/A}) due to a limited convergence ceiling. As the ratio of slow clients increases, efficiency dynamically shifts toward smaller $M$ values, exposing a sharp performance drop whenever $M$ exceeds the fast-client count ($M > N - N_{\text{slow}}$) and forces server stalls.
\end{itemize}

Importantly, these findings remain robust as data heterogeneity increases, demonstrating that the semi-asynchronous scheduling strategy effectively adapts to diverse system and statistical bottlenecks.

\begin{table*}[t]
\centering
\small
\setlength{\tabcolsep}{4pt}
\renewcommand{\arraystretch}{0.85}
\noindent\textbf{Slow = 0\%} \\[0.2em]
\resizebox{\textwidth}{!}{%
\begin{tabular}{lccccccc}
\hline
Strategy & $\Delta_{loss}/time$ & Final Acc. & Max Acc. & TTA 50\% & TTA 58\% & Acc@15s & Acc@25s \\
\hline
FedSaSync ($M = 8$) & \textbf{0.0204 $\pm$ 0.0004} & 52.5 $\pm$ 0.9\% & 52.5 $\pm$ 0.9\% & 37.3 $\pm$ 1.8s & N/A & 39.5 $\pm$ 1.1\% & 45.2 $\pm$ 0.6\% \\
FedSaSync ($M = 10$) & 0.0157 $\pm$ 0.0006 & 55.6 $\pm$ 0.9\% & 55.6 $\pm$ 0.9\% & 36.8 $\pm$ 1.9s & N/A & 39.8 $\pm$ 0.7\% & 45.3 $\pm$ 0.6\% \\
FedSaSync ($M = 12$) & 0.0136 $\pm$ 0.0004 & 57.8 $\pm$ 1.6\% & 57.8 $\pm$ 1.6\% & 36.8 $\pm$ 1.9s & 64.0 $\pm$ 2.4s\textsuperscript{3} & 40.3 $\pm$ 0.7\% & 45.7 $\pm$ 0.4\% \\
FedSaSync ($M = 14$) & 0.0124 $\pm$ 0.0004 & 58.8 $\pm$ 1.8\% & 58.8 $\pm$ 1.8\% & 35.6 $\pm$ 1.6s & 66.7 $\pm$ 2.9s\textsuperscript{4} & 40.6 $\pm$ 0.4\% & 45.9 $\pm$ 0.6\% \\
FedSaSync ($M = 16$) & 0.0109 $\pm$ 0.0006 & 60.1 $\pm$ 1.9\% & 60.1 $\pm$ 1.9\% & \textbf{33.9 $\pm$ 1.2s} & \textbf{63.3 $\pm$ 3.2s\textsuperscript{4}} & 40.5 $\pm$ 0.7\% & 46.0 $\pm$ 0.4\% \\
FedSaSync ($M = 18$) & 0.0097 $\pm$ 0.0004 & 61.4 $\pm$ 2.1\% & 61.4 $\pm$ 2.1\% & 34.4 $\pm$ 0.7s & 63.5 $\pm$ 3.5s\textsuperscript{4} & 40.4 $\pm$ 0.6\% & \textbf{46.1 $\pm$ 0.3\%} \\
FedSaSync ($M = 20$) & 0.0093 $\pm$ 0.0005 & 61.8 $\pm$ 2.4\% & 61.8 $\pm$ 2.4\% & 35.0 $\pm$ 1.4s & 63.4 $\pm$ 3.3s\textsuperscript{4} & \textbf{40.8 $\pm$ 0.5\%} & 46.1 $\pm$ 0.4\% \\
FedAvg & 0.0094 $\pm$ 0.0005 & \textbf{61.9 $\pm$ 2.2\%} & \textbf{61.9 $\pm$ 2.2\%} & 34.4 $\pm$ 1.5s & 68.3 $\pm$ 10.2s & 40.2 $\pm$ 0.5\% & \textbf{46.1 $\pm$ 0.3\%} \\
\hline
\end{tabular}%
}
\par\vspace{0.8em}
\noindent\textbf{Slow = 20\%} \\[0.2em]
\resizebox{\textwidth}{!}{%
\begin{tabular}{lccccccc}
\hline
Strategy & $\Delta_{loss}/time$ & Final Acc. & Max Acc. & TTA 50\% & TTA 58\% & Acc@15s & Acc@25s \\
\hline
FedSaSync ($M = 8$) & \textbf{0.0143 $\pm$ 0.0032} & 52.9 $\pm$ 1.6\% & 52.9 $\pm$ 1.6\% & 37.9 $\pm$ 4.3s & N/A & 39.3 $\pm$ 0.9\% & 45.2 $\pm$ 1.4\% \\
FedSaSync ($M = 10$) & 0.0118 $\pm$ 0.0007 & 56.3 $\pm$ 1.5\% & 56.3 $\pm$ 1.5\% & 36.8 $\pm$ 2.6s & N/A & \textbf{41.0 $\pm$ 0.9\%} & 46.0 $\pm$ 0.6\% \\
FedSaSync ($M = 12$) & 0.0118 $\pm$ 0.0003 & 57.2 $\pm$ 1.9\% & 57.2 $\pm$ 1.9\% & 36.1 $\pm$ 2.3s & N/A\textsuperscript{2} & 40.6 $\pm$ 0.6\% & 45.9 $\pm$ 0.4\% \\
FedSaSync ($M = 14$) & 0.0107 $\pm$ 0.0003 & 58.8 $\pm$ 1.9\% & 58.8 $\pm$ 1.9\% & \textbf{35.9 $\pm$ 2.6s} & 67.8 $\pm$ 3.5s\textsuperscript{4} & 39.9 $\pm$ 0.5\% & 46.0 $\pm$ 0.6\% \\
FedSaSync ($M = 16$) & 0.0103 $\pm$ 0.0005 & 59.9 $\pm$ 2.1\% & 59.9 $\pm$ 2.1\% & 36.1 $\pm$ 2.7s & \textbf{65.8 $\pm$ 3.1s\textsuperscript{4}} & 40.3 $\pm$ 1.1\% & \textbf{46.0 $\pm$ 0.3\%} \\
FedSaSync ($M = 18$) & 0.0062 $\pm$ 0.0003 & 60.7 $\pm$ 1.9\% & 60.7 $\pm$ 1.9\% & 64.7 $\pm$ 3.5s & 116.2 $\pm$ 6.2s\textsuperscript{4} & 32.6 $\pm$ 3.0\% & 39.2 $\pm$ 1.8\% \\
FedSaSync ($M = 20$) & 0.0033 $\pm$ 0.0002 & \textbf{61.9 $\pm$ 2.4\%} & \textbf{61.9 $\pm$ 2.4\%} & 114.4 $\pm$ 4.5s & 208.4 $\pm$ 9.0s\textsuperscript{4} & 25.6 $\pm$ 1.3\% & 32.7 $\pm$ 0.6\% \\
FedAvg & 0.0033 $\pm$ 0.0002 & 61.7 $\pm$ 2.3\% & 61.7 $\pm$ 2.3\% & 115.8 $\pm$ 3.8s & 208.7 $\pm$ 9.0s\textsuperscript{4} & 25.6 $\pm$ 1.3\% & 32.7 $\pm$ 0.5\% \\
\hline
\end{tabular}%
}
\par\vspace{0.8em}
\noindent\textbf{Slow = 50\%} \\[0.2em]
\resizebox{\textwidth}{!}{%
\begin{tabular}{lccccccc}
\hline
Strategy & $\Delta_{loss}/time$ & Final Acc. & Max Acc. & TTA 50\% & TTA 58\% & Acc@15s & Acc@25s \\
\hline
FedSaSync ($M = 8$) & \textbf{0.0152 $\pm$ 0.0003} & 54.0 $\pm$ 1.1\% & 54.0 $\pm$ 1.1\% & \textbf{36.8 $\pm$ 1.7s} & N/A & \textbf{41.0 $\pm$ 0.4\%} & \textbf{46.1 $\pm$ 0.6\%} \\
FedSaSync ($M = 10$) & 0.0130 $\pm$ 0.0004 & 55.7 $\pm$ 1.5\% & 55.7 $\pm$ 1.5\% & 39.7 $\pm$ 4.0s & N/A & 39.1 $\pm$ 0.7\% & 45.3 $\pm$ 0.4\% \\
FedSaSync ($M = 12$) & 0.0097 $\pm$ 0.0004 & 56.6 $\pm$ 1.4\% & 56.6 $\pm$ 1.4\% & 54.8 $\pm$ 4.6s & N/A & 33.0 $\pm$ 1.3\% & 40.1 $\pm$ 1.2\% \\
FedSaSync ($M = 14$) & 0.0064 $\pm$ 0.0002 & 58.6 $\pm$ 1.7\% & 58.6 $\pm$ 1.7\% & 76.1 $\pm$ 3.0s & \textbf{139.4 $\pm$ 5.2s\textsuperscript{4}} & 32.4 $\pm$ 0.7\% & 37.8 $\pm$ 0.8\% \\
FedSaSync ($M = 16$) & 0.0034 $\pm$ 0.0001 & 60.8 $\pm$ 2.1\% & 60.8 $\pm$ 2.1\% & 124.3 $\pm$ 6.0s & 226.1 $\pm$ 15.0s\textsuperscript{4} & 25.4 $\pm$ 1.6\% & 32.2 $\pm$ 0.6\% \\
FedSaSync ($M = 18$) & 0.0033 $\pm$ 0.0001 & 61.2 $\pm$ 2.1\% & 61.2 $\pm$ 2.1\% & 122.5 $\pm$ 7.2s & 219.7 $\pm$ 9.2s\textsuperscript{4} & 25.4 $\pm$ 1.3\% & 32.4 $\pm$ 0.6\% \\
FedSaSync ($M = 20$) & 0.0031 $\pm$ 0.0001 & \textbf{61.7 $\pm$ 2.2\%} & \textbf{61.7 $\pm$ 2.2\%} & 121.7 $\pm$ 4.9s & 221.8 $\pm$ 9.6s\textsuperscript{4} & 25.6 $\pm$ 1.3\% & 29.6 $\pm$ 0.7\% \\
FedAvg & 0.0031 $\pm$ 0.0001 & 61.7 $\pm$ 2.4\% & 61.7 $\pm$ 2.4\% & 123.3 $\pm$ 4.2s & 220.5 $\pm$ 10.1s\textsuperscript{4} & 25.6 $\pm$ 1.3\% & 29.6 $\pm$ 0.8\% \\
\hline
\end{tabular}%
}
\par\vspace{0.8em}
\caption{Performance and efficiency metrics for CIFAR10 (IID) under different semi-asynchronous configurations and fractions of clients.}
\label{tab:comprehensive_cifar10_iid}
\raggedright \footnotesize \textit{Note:} Superscripts indicate the number of successful runs out of 5 repetitions when not all runs finished within the target. Every result obtained from 2 or fewer successful runs is labeled as N/A.
\end{table*}

\subsubsection{Fairness and Staleness Analysis}
Beyond convergence speed, it is crucial to analyze fairness and staleness to understand the underlying behavioral dynamics of the system. Because execution times vary, unmonitored semi-asynchronous strategies can potentially introduce client starvation or accumulate high staleness when slow clients batch together. Evaluating these metrics ensures that the strategy operates correctly and provides visibility into otherwise hidden behavioral patterns.

\begin{figure*}[t]
    \centering
    \includegraphics[width=1.0\linewidth]{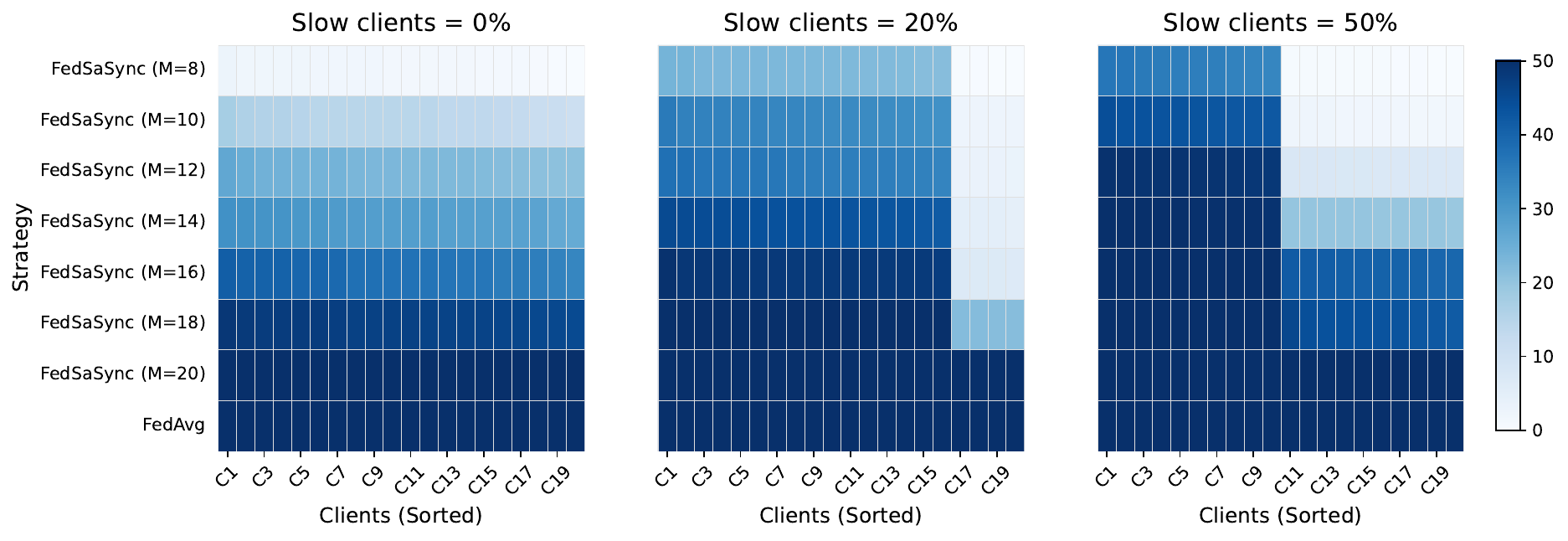}
    \caption{Client fairness for CIFAR-10 (IID) under different semi-asynchronous configurations and fractions of slow clients.}
    \label{fig:cifar_comparison_fairness}
\end{figure*}

\begin{figure*}[t]
    \centering
    \includegraphics[width=1.0\linewidth]{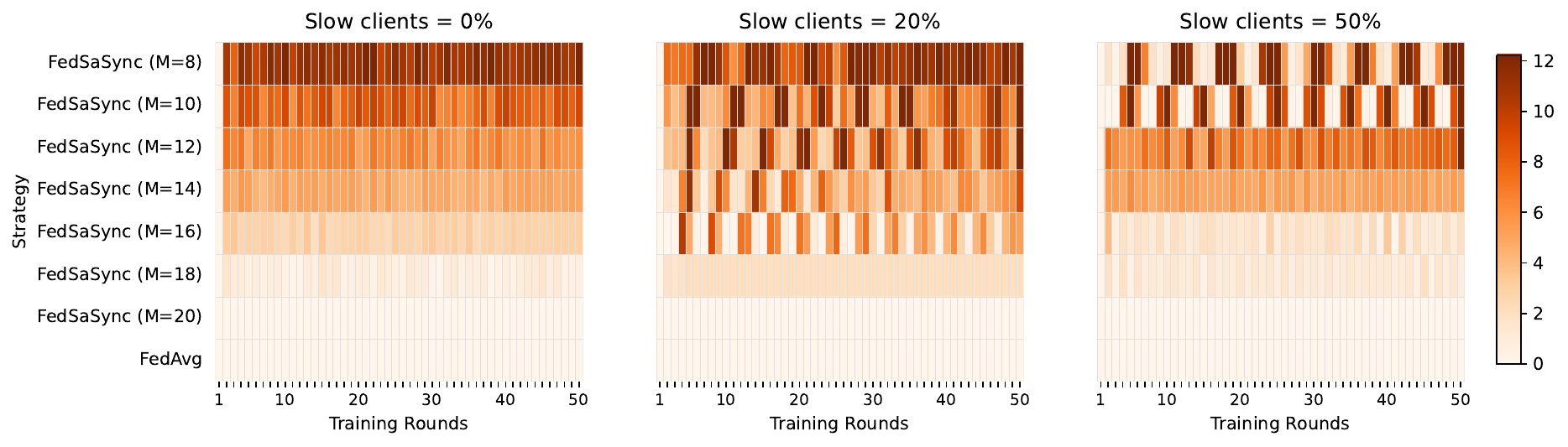}
    \caption{Round accumulated staleness for CIFAR-10 (IID) under different semi-asynchronous configurations and fractions of slow clients.}
    \label{fig:cifar_comparison_staleness}
\end{figure*}

Figure~\ref{fig:cifar_comparison_fairness} illustrates client participation fairness. Under IID data distribution, groups of clients (both fast and slow) generally maintain balanced participation levels because the system naturally clusters them as they complete local workloads within similar windows. However, configurations with lower $M$ thresholds reduce overall client participation—since fewer updates are required to trigger aggregation—which directly accounts for the system's enhanced speed. As the ratio of slow clients increases, the critical horizon of $M = N - N_{\text{slow}}$ becomes clearly distinguishable, mirroring the trends observed in our loss and accuracy evaluations. Specifically, when thresholds cross this boundary, stragglers are systematically bypassed, causing a significant drop in their participation frequency.

To complement this, Figure~\ref{fig:cifar_comparison_staleness} presents the round-accumulated staleness over 50 rounds. Staleness and fairness are deeply intertwined: larger $M$ configurations inherently reduce accumulated staleness by waiting for a broader pool, preventing updates from becoming outdated. Nevertheless, all executions exhibit a distinct periodicity where slow clients' execution times periodically align across certain rounds. Interestingly, even with periodic staleness spikes in lower $M$ configurations (especially under a 50\% slow client ratio), the underlying staleness within the critical horizon remains remarkably low. This confirms that the scheduler successfully captures and adapts to underlying client profiles.

\subsubsection{Scheduler Overhead}

Finally, Table~\ref{tab:scheduler_cifar10_iid} reports the scheduler and communication overhead metrics, providing insight into the infrastructure cost of the semi-asynchronous framework.

\begin{itemize}[leftmargin=1.2em, nosep]
    \item \textbf{Polling Calls:} Configurations with lower $M$ thresholds consistently incur fewer total polling calls, as faster-completing rounds bypass stragglers and reduce the frequency of server checks.
    \item \textbf{Polling Overhead (\%):} Despite varying call counts, relative polling overhead remains minimal and stable ($\approx 0.4\% - 0.6\%$), as the asynchronous mechanism avoids network serialization bottlenecks regardless of frequency.
    \item \textbf{Server CPU Utilization (\%):} Lower $M$ values more than double server CPU usage compared to synchronous baselines (reaching $23.4\%$ for $M = 8$ at $\text{Slow} = 0.0$). Because smaller thresholds trigger aggregation tasks at a much higher frequency, the server spends more time actively computing rather than idling.
    \item \textbf{Aggregation Rate (msgs/s):} Lower $M$ settings achieve peak throughput when uninterrupted (e.g., $14.16\text{ msgs/s}$ for $M = 8$ at $\text{Slow} = 0.0$ by continuously ingesting fast updates). As straggler delays increase, intermediate configurations aligned with non-straggler capacity maintain superior message throughput over synchronous baselines.
\end{itemize}

\begin{table*}[t]
\centering
\small
\setlength{\tabcolsep}{4pt}
\renewcommand{\arraystretch}{0.85}
\noindent\textbf{Slow = 0\%} \\[0.2em]
\resizebox{\textwidth}{!}{%
\begin{tabular}{lcccc}
\hline
Strategy & Polling Calls & Polling Overhead (\%) & Server CPU Util. (\%) & Agg. Rate (msgs/s) \\
\hline
FedSaSync ($M = 8$) & \textbf{663.4 $\pm$ 18.5} & \textbf{0.4 $\pm$ 0.0\%} & 23.4 $\pm$ 0.7\% & \textbf{14.16 $\pm$ 0.44} \\
FedSaSync ($M = 10$) & 1040.8 $\pm$ 103.4 & \textbf{0.4 $\pm$ 0.1\%} & 19.2 $\pm$ 2.4\% & 11.67 $\pm$ 1.35 \\
FedSaSync ($M = 12$) & 1291.0 $\pm$ 18.5 & 0.5 $\pm$ 0.0\% & 18.4 $\pm$ 0.4\% & 10.90 $\pm$ 0.18 \\
FedSaSync ($M = 14$) & 1468.2 $\pm$ 29.3 & 0.5 $\pm$ 0.0\% & 17.8 $\pm$ 0.5\% & 10.47 $\pm$ 0.25 \\
FedSaSync ($M = 16$) & 1674.2 $\pm$ 41.0 & 0.5 $\pm$ 0.0\% & 17.4 $\pm$ 0.5\% & 10.30 $\pm$ 0.26 \\
FedSaSync ($M = 18$) & 1854.6 $\pm$ 10.4 & 0.5 $\pm$ 0.0\% & 17.1 $\pm$ 0.2\% & 10.38 $\pm$ 0.11 \\
FedSaSync ($M = 20$) & 1912.4 $\pm$ 10.5 & 0.6 $\pm$ 0.0\% & \textbf{17.0 $\pm$ 0.2\%} & 10.44 $\pm$ 0.07 \\
FedAvg & 1887.4 $\pm$ 14.1 & 0.6 $\pm$ 0.0\% & 17.1 $\pm$ 0.2\% & 10.59 $\pm$ 0.08 \\
\hline
\end{tabular}%
}
\par\vspace{0.8em}
\noindent\textbf{Slow = 20\%} \\[0.2em]
\resizebox{\textwidth}{!}{%
\begin{tabular}{lcccc}
\hline
Strategy & Polling Calls & Polling Overhead (\%) & Server CPU Util. (\%) & Agg. Rate (msgs/s) \\
\hline
FedSaSync ($M = 8$) & \textbf{1310.8 $\pm$ 479.3} & 0.6 $\pm$ 0.4\% & 14.1 $\pm$ 5.8\% & 6.92 $\pm$ 2.64 \\
FedSaSync ($M = 10$) & 1631.0 $\pm$ 209.7 & 0.5 $\pm$ 0.1\% & 14.5 $\pm$ 2.1\% & 7.45 $\pm$ 1.01 \\
FedSaSync ($M = 12$) & 1594.8 $\pm$ 19.5 & 0.5 $\pm$ 0.0\% & 14.8 $\pm$ 0.4\% & 8.11 $\pm$ 0.13 \\
FedSaSync ($M = 14$) & 1745.2 $\pm$ 97.0 & 0.5 $\pm$ 0.1\% & 15.2 $\pm$ 1.0\% & 8.69 $\pm$ 0.50 \\
FedSaSync ($M = 16$) & 1769.2 $\pm$ 14.8 & 0.5 $\pm$ 0.0\% & 15.4 $\pm$ 0.4\% & \textbf{9.36 $\pm$ 0.11} \\
FedSaSync ($M = 18$) & 3180.6 $\pm$ 79.9 & \textbf{0.4 $\pm$ 0.0\%} & 10.2 $\pm$ 0.3\% & 5.67 $\pm$ 0.15 \\
FedSaSync ($M = 20$) & 6066.0 $\pm$ 5.1 & \textbf{0.4 $\pm$ 0.0\%} & \textbf{6.7 $\pm$ 0.1\%} & 3.28 $\pm$ 0.00 \\
FedAvg & 6079.6 $\pm$ 5.0 & \textbf{0.4 $\pm$ 0.0\%} & \textbf{6.7 $\pm$ 0.0\%} & 3.27 $\pm$ 0.00 \\
\hline
\end{tabular}%
}
\par\vspace{0.8em}
\noindent\textbf{Slow = 50\%} \\[0.2em]
\resizebox{\textwidth}{!}{%
\begin{tabular}{lcccc}
\hline
Strategy & Polling Calls & Polling Overhead (\%) & Server CPU Util. (\%) & Agg. Rate (msgs/s) \\
\hline
FedSaSync ($M = 8$) & \textbf{1205.6 $\pm$ 25.5} & 0.5 $\pm$ 0.0\% & 16.0 $\pm$ 0.5\% & \textbf{7.79 $\pm$ 0.23} \\
FedSaSync ($M = 10$) & 1409.8 $\pm$ 46.9 & \textbf{0.4 $\pm$ 0.0\%} & 15.3 $\pm$ 0.6\% & 7.78 $\pm$ 0.28 \\
FedSaSync ($M = 12$) & 2072.2 $\pm$ 101.2 & 0.5 $\pm$ 0.0\% & 12.6 $\pm$ 0.7\% & 6.17 $\pm$ 0.32 \\
FedSaSync ($M = 14$) & 3201.8 $\pm$ 29.0 & 0.5 $\pm$ 0.0\% & 9.8 $\pm$ 0.1\% & 4.64 $\pm$ 0.05 \\
FedSaSync ($M = 16$) & 6076.2 $\pm$ 40.0 & 0.5 $\pm$ 0.0\% & \textbf{7.1 $\pm$ 0.1\%} & 2.99 $\pm$ 0.03 \\
FedSaSync ($M = 18$) & 6136.2 $\pm$ 9.8 & 0.5 $\pm$ 0.0\% & 7.2 $\pm$ 0.0\% & 3.04 $\pm$ 0.01 \\
FedSaSync ($M = 20$) & 6389.4 $\pm$ 15.7 & 0.5 $\pm$ 0.0\% & 7.4 $\pm$ 0.0\% & 3.09 $\pm$ 0.01 \\
FedAvg & 6406.6 $\pm$ 10.8 & 0.5 $\pm$ 0.0\% & 7.4 $\pm$ 0.0\% & 3.08 $\pm$ 0.01 \\
\hline
\end{tabular}%
}
\caption{Scheduler and communication overhead metrics for CIFAR10 (IID) under different semi-asynchronous configurations and fractions of slow clients.}
\label{tab:scheduler_cifar10_iid}
\end{table*}

A similar trend is observed for MNIST (see GitHub repository).

\subsubsection{Impact of Data Heterogeneity}
To validate the robustness of our framework under statistical heterogeneity, we analyze its performance across non-IID data distributions, where the local datasets are partitioned using a Dirichlet distribution.

\begin{figure*}[!htbp]
    \centering
    \includegraphics[width=1.0\linewidth]{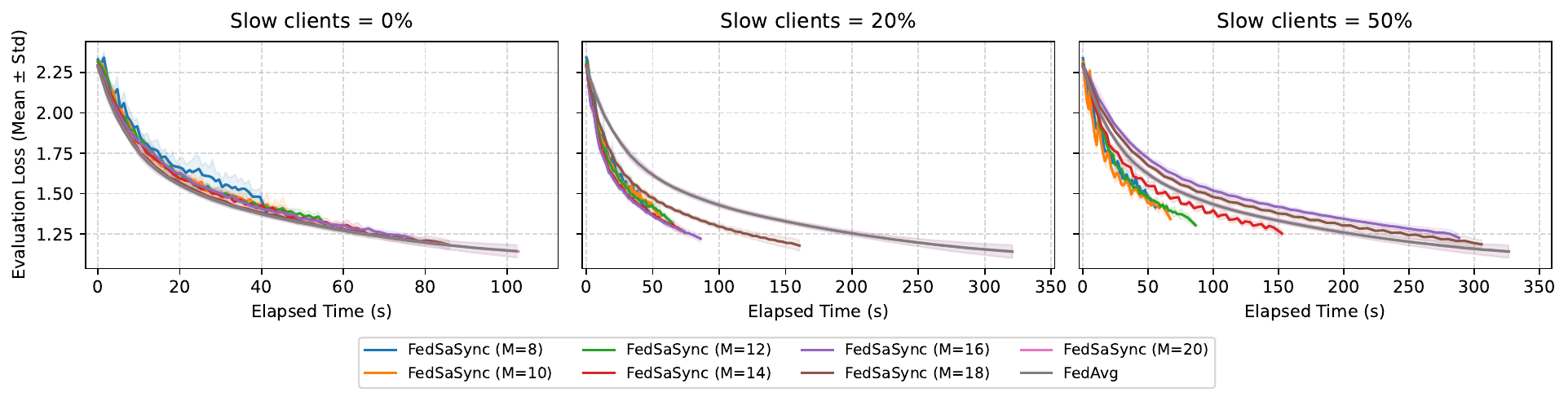}
    \caption{Evaluation loss versus wall-clock time for CIFAR-10 (non-IID $\alpha = 0.5$) under different semi-asynchronous configurations and fractions of slow clients.}
    \label{fig:cifar_noniid_loss}
\end{figure*}

\textbf{Convergence Dynamics ($\text{Loss}$ vs. $\text{Time}$):} 
As expected, non-IID loss curves in Figure~\ref{fig:cifar_noniid_loss} exhibit greater instability due to client drift, yet follow the same overarching operational trends observed in the IID experiments. Notably, configurations with overly high $M$ thresholds perform worse than standard \texttt{FedAvg} under non-IID conditions, as waiting for a large quorum of skewed updates amplifies gradient conflicts and delays global progress.

\begin{figure*}[!htbp]
    \centering
    \includegraphics[width=1.0\linewidth]{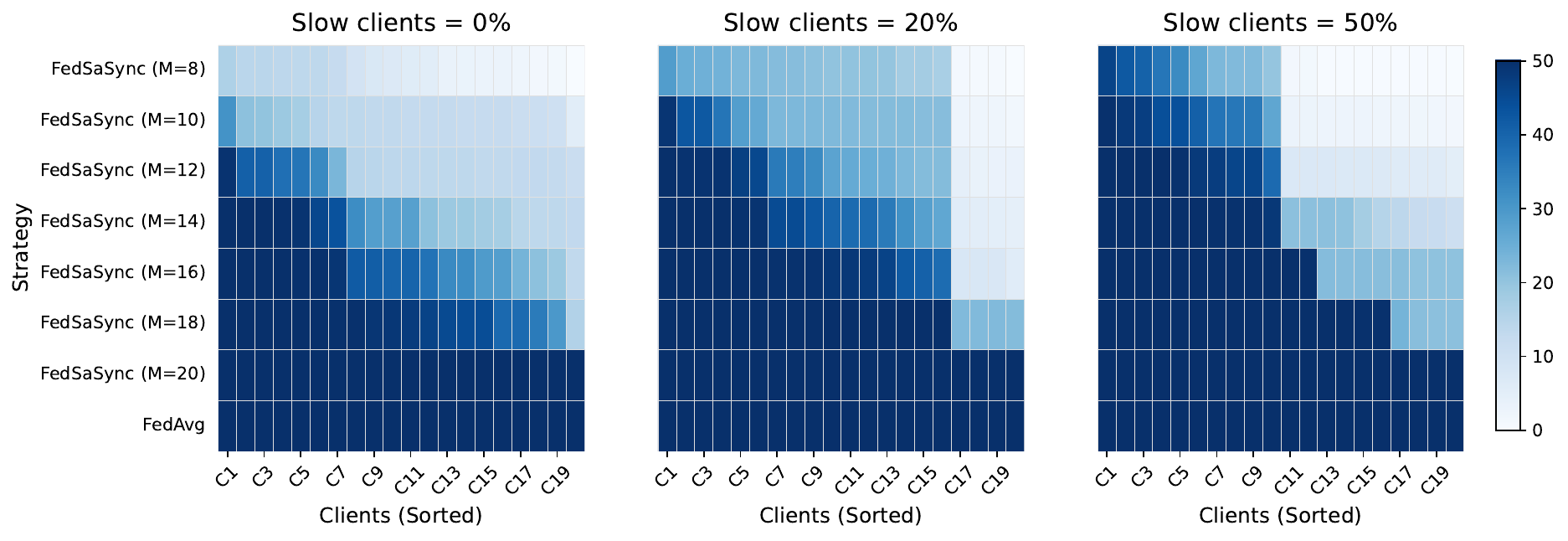}
    \caption{Client fairness for CIFAR-10 (non-IID $\alpha = 0.5$) under different semi-asynchronous configurations and fractions of slow clients.}
    \label{fig:cifar_noniid_comparison_fairness}
\end{figure*}

\textbf{Client Participation Fairness:} 
The fairness dynamics in Figure~\ref{fig:cifar_noniid_comparison_fairness} reveal distinct behavioral shifts. Under IID distributions, clients neatly synchronize because they hold identical data volumes and finish training within aligned windows. In contrast, non-IID partitioning shatters this uniformity: client participation becomes ``blurred'' and dispersed, as nodes with smaller local datasets complete workloads faster and contribute more frequently. Nevertheless, the underlying separation in participation trends between fast and slow clients remains robustly preserved.

\begin{figure*}[!htbp]
    \centering
    \includegraphics[width=1.0\linewidth]{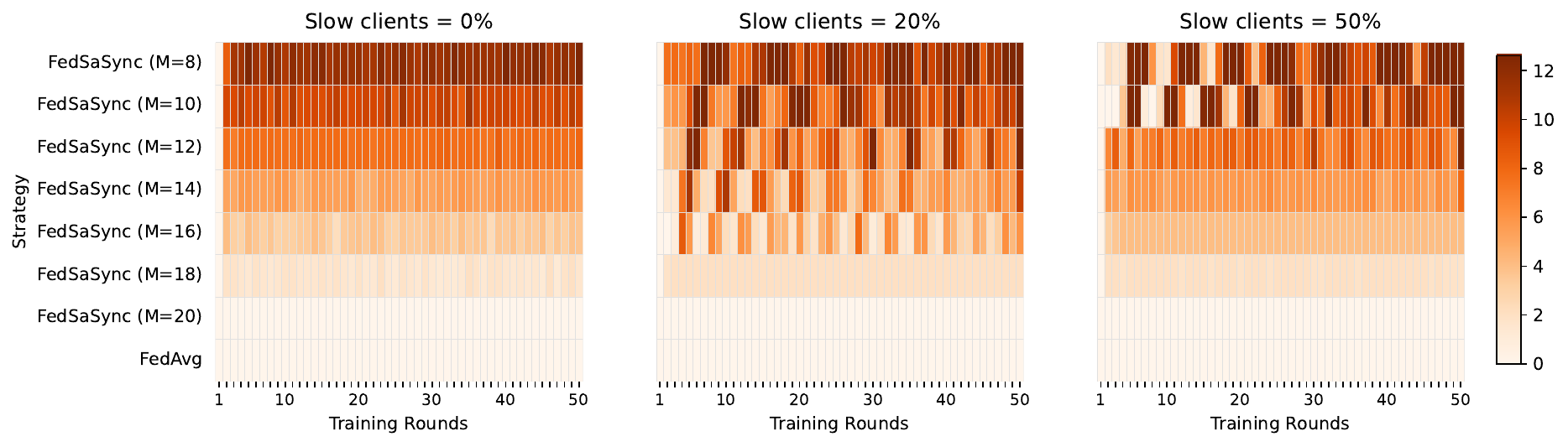}
    \caption{Round accumulated staleness for CIFAR-10 (non-IID $\alpha = 0.5$) under different semi-asynchronous configurations and fractions of slow clients.}
    \label{fig:cifar_noniid_comparison_staleness}
\end{figure*}

\textbf{Accumulated Staleness:} 
Staleness patterns under non-IID settings are notably more distributed across rounds in Figure~\ref{fig:cifar_noniid_comparison_staleness}. The natural asynchrony and variance in local dataset sizes prevent clients from clustering strictly, which mitigates severe periodic staleness spikes. Consequently, while the underlying baseline staleness remains slightly higher for lower $M$ configurations due to frequent updates from fast nodes, the absolute peak staleness values across rounds are reduced compared to the IID scenario.

These results demonstrate the viability of the proposed implementation and confirm that the semi-asynchronous mechanism operates as intended. Overall, the framework provides a solid foundation for further investigation of \ac{SAFL} under heterogeneous environments.

\subsubsection{Scalability Analysis}
\label{sec:scalability}
To validate the scalability of the system, we performed cross-silo experiments scaling up to 100 clients under simulated execution times. The Ray backend framework was deployed across a wired local subnet comprising three computation nodes: two identical nodes equipped with dual-socket Intel Xeon E5-2650 v2 processors (contributing 30 logical CPUs each) and a third distinct node featuring Intel Xeon E5-2650 v4 processors (contributing 40 logical CPUs), summing up to exactly 100 logical CPUs to satisfy the resource allocation constraint of one client per logical CPU.

To ensure clearly discernible trends in the results, we adjusted specific hyperparameters; notably, local epochs were increased from 1 to 3 per client to compensate for the data sparsity associated with a larger client pool. Furthermore, due to severe CPU compute constraints, the evaluation phase was migrated from a distributed to a centralized process (i.e., \texttt{fraction-evaluate = 0.0}). This prevents slower straggler clients from bottlenecking the system by overlapping local training with concurrent resource-heavy evaluations, which would otherwise dominate wall-clock time due to resource contention.

\begin{figure*}[!htbp]
\centering
\includegraphics[width=0.9\linewidth]{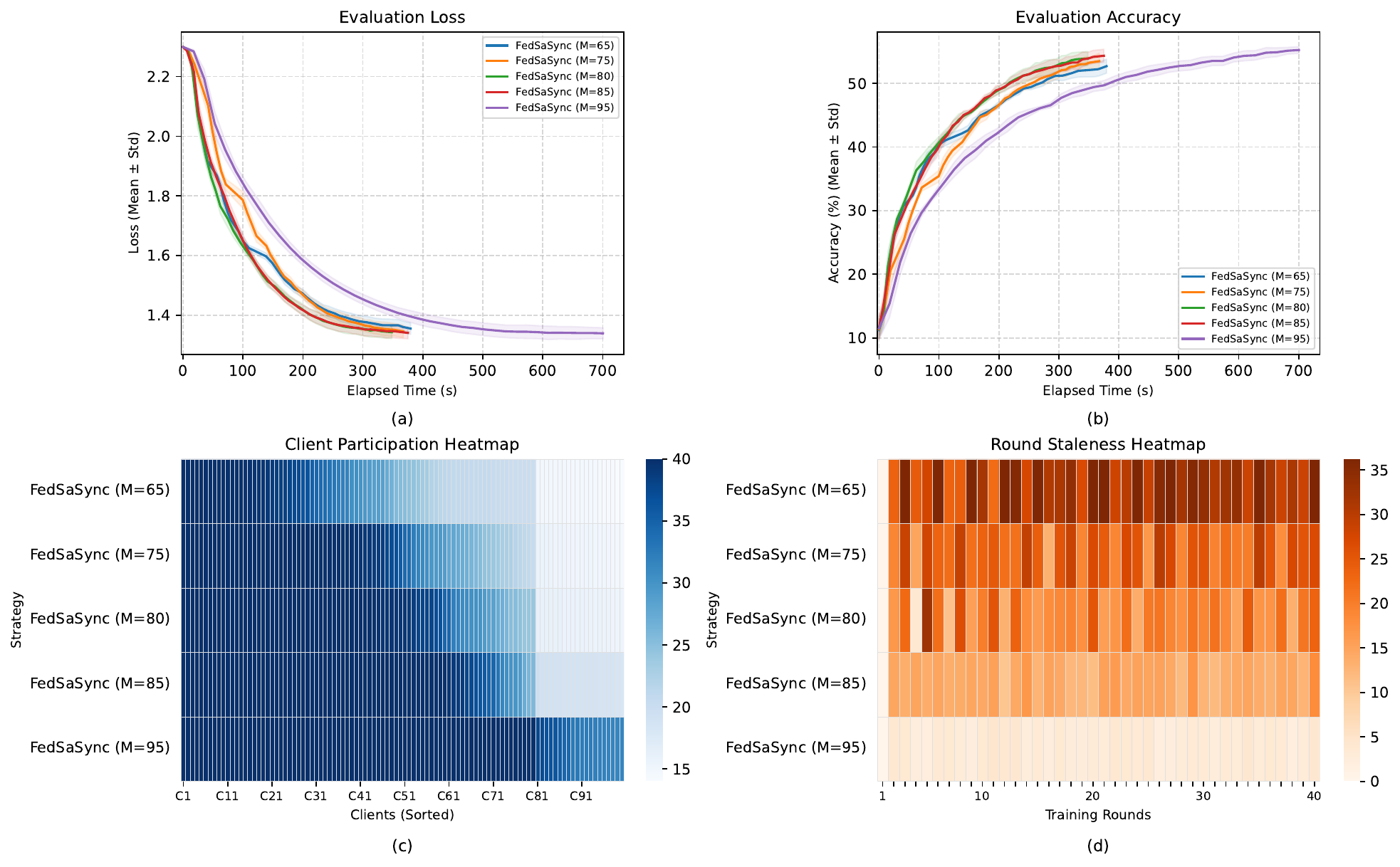}
\caption{Performance and behavior metrics grid for CIFAR-10 (IID) under 20\% slow clients with 100 participants: (a) evaluation loss over elapsed time, (b) evaluation accuracy over elapsed time, (c) client participation heatmap across sorted nodes, and (d) round staleness heatmap progression across training rounds.}
\label{fig:metrics_grid_cifar10_iid}
\end{figure*}

Figure~\ref{fig:metrics_grid_cifar10_iid} illustrates this scalability analysis. The subplots demonstrate that the system preserves the operational structure and convergence behavior observed in smaller-scale deployments, thereby confirming its robust scalability in cross-silo environments (additional plots and execution instructions are available in the public repository).

\section{Software limitations}
The proposed implementation is built on top of the Flower framework, and therefore inherits certain design choices and constraints specific to this environment. In particular, Flower relies on a \textit{Grid} abstraction to manage federated clients and coordinate communication between them. As a consequence, the presented semi-asynchronous implementation is partially coupled to the framework's underlying execution model. While the general concept of semi-asynchronous coordination is not framework-dependent, its concrete realization may vary across different \ac{FL} systems.

In addition, the current implementation is subject to limitations inherent in the communication backend's design. The \textit{Grid} component is primarily optimized for synchronous execution patterns, in which coordinated broadcast and aggregation operations dominate the communication paradigm. As a result, introducing semi-asynchronous behavior may lead to suboptimal performance under certain conditions, particularly in scenarios involving large numbers of clients or highly variable communication patterns.

From an experimental standpoint, wall-clock time evaluation on a single workstation is constrained by hardware concurrency. Flower relies on Ray to schedule client execution across available CPU resource slots (operating independently of low-level OS core affinity). When the number of concurrent simulated clients exceeds the machine's logical cores, Ray queues the remaining tasks. In heterogeneous settings where delayed clients hold CPU slots, faster clients stall in the scheduling queue, introducing artificial head-of-line blocking. Under these conditions, wall-clock measurements reflect CPU contention rather than true protocol dynamics. Consequently, bounding execution concurrency to match the available logical cores was adopted as a controlled baseline configuration for our testbed to preserve wall-clock fidelity, whereas evaluating larger client populations without temporal distortion can also be addressed by expanding the Ray cluster across multiple physical nodes (as shown in Section~\ref{sec:scalability}).

Furthermore, the proposed approach requires the semi-asynchronous degree to be defined a priori and remain fixed throughout the training process. This constitutes a significant limitation, as it assumes some knowledge of client performance characteristics before deployment. In real-world federated environments, particularly in \ac{IoT} scenarios, client availability, computational capacity, and communication quality may vary dynamically over time, while unexpected communication failures can further alter system behavior. Consequently, a static semi-asynchronous degree may become suboptimal during execution, leading to inefficient synchronization decisions and limiting the framework's ability to adapt to changing operating conditions.

\section{Impact}
This work extends the Flower framework with support for an SAFL mechanism, demonstrating how semi-asynchronous federated learning can be incorporated into an existing federated learning platform. Although the implementation has been developed with compatibility and reusability in mind, adoption within the broader framework ecosystem would require additional validation, maintenance efforts, and community acceptance.

Nevertheless, the proposed implementation constitutes a useful research artifact that facilitates the evaluation of semi-asynchronous execution strategies in realistic federated learning environments. By making the source code publicly available, the work aims to support reproducibility and provide a foundation for future research and potential framework-level extensions.

\section{Conclusions}
This work proposes FedSaSync, an extension of the Flower framework to implement \ac{SAFL}, enabling federated systems to better adapt to heterogeneous and low-resource environments, achieving a middle-ground between fully synchronous \ac{FL} and asynchronous \ac{FL}. The software architecture and operational mechanisms are described in detail, providing a comprehensive overview of the proposed mechanism.

The proposed strategy has been tested in simulated environments with time-varying clients to model heterogeneous conditions. Experiments were conducted on the CIFAR10 and MNIST datasets, varying the simulated slow clients to analyze heterogeneous environments of different scales. The experimental results show that the FedSaSync strategy behaves similarly to FedAvg when all clients are homogeneous, and outperforms FedAvg as system heterogeneity increases, demonstrating that semi-asynchronous coordination can effectively mitigate the impact of client heterogeneity by reducing idle time and improving resource utilization.

Despite the promising results, the proposed approach remains subject to several limitations, including its dependence on the Flower execution model, the constraints imposed by a communication backend optimized for synchronous operation, and the use of a fixed semi-asynchronous degree. These factors may limit scalability and adaptability in highly dynamic federated environments, motivating future research on framework-independent and adaptive semi-asynchronous strategies. In particular, the development of infrastructure-agnostic semi-asynchronous strategies, able to empirically model the behavior of clients by leveraging metadata inherently present in standard FL workflows, without requiring additional client profiling or telemetry to avoid overheads, will be explored as future work.

Overall, the results indicate that FedSaSync constitutes a promising initial approach to \ac{SAFL} in heterogeneous environments and lays the foundation for future research on adaptive coordination mechanisms.




\section*{Acknowledgements}
\label{sec:acknowledgements}
This publication is part of the I+D+i project PID2024-158682OB-C32, funded by MICIU/AEI/10.13039/501100011033/and FEDER/UE.
It has also been supported by 2025-GRIN-38312 grant funded by the University of Castilla-La Mancha for Consolidated Research Groups. 
The first author is also supported by a predoctoral contract funded by the University of Castilla-La Mancha (UCLM) and co-financed by the European Union through the European Social Fund Plus (ESF+).

\section*{Declaration of generative AI and AI-assisted technologies in the manuscript preparation process}
\label{sec:ai_declaration}

During the preparation of this work, the author(s) used Grammarly, ChatGPT (OpenAI), and Microsoft Copilot in order to improve language, clarity, and readability. The use of these tools was limited to language refinement and information support. After using these tools, the authors carefully reviewed, edited, and validated the content as needed, and take full responsibility for the final content of the publication.










\begin{acronym}
    \acro{FL}{Federated Learning}
    \acro{SAFL}{Semi-Asynchronous Federated Learning}
    \acro{AFL}{Asynchronous Federated Learning}
    \acro{AI}{Artificial Intelligence}
    \acro{ML}{Machine Learning}
    \acro{DL}{Deep Learning}
    \acro{IoT}{Internet of Things}
    \acro{EC}{Edge Computing}
    \acro{FC}{Fog Computing}
    \acro{CC}{Cloud Computing}
\end{acronym}

\end{document}